\documentclass[11pt]{article}

 \usepackage{amssymb}
 \usepackage{amsmath}
 \usepackage{amsfonts}
 \usepackage{amsthm}
 \usepackage{pifont}    %% to get \ding{} commands used in \mybullet
 \usepackage{stmaryrd}  %% for \llceil, \rrceil and \Yright commands
 \usepackage{url}
 \usepackage{float}
 \usepackage{graphicx}
 \usepackage{cite}
 \usepackage{fancybox}
 \usepackage{varioref}
 \usepackage{xspace}

\ifx\pdftexversion\undefined
  \usepackage[colorlinks,linkcolor=black,filecolor=black,citecolor=black,urlcolor=black,pdfstartview=FitH]{hyperref}
\else
  \usepackage[colorlinks,linkcolor=blue,filecolor=blue,citecolor=blue,urlcolor=blue,pdfstartview=FitH]{hyperref}
\fi

\newsavebox{\theorembox}
\newsavebox{\lemmabox}
\newsavebox{\conjecturebox}
\newsavebox{\claimbox}
\newsavebox{\factbox}
\newsavebox{\corollarybox}
\newsavebox{\propositionbox}
\newsavebox{\examplebox}
\savebox{\theorembox}{\bf Theorem}
\savebox{\lemmabox}{\bf Lemma}
\savebox{\conjecturebox}{\bf Conjecture}
\savebox{\claimbox}{\bf Claim}
\savebox{\factbox}{\bf Fact}
\savebox{\corollarybox}{\bf Corollary}
\savebox{\propositionbox}{\bf Proposition}
\savebox{\examplebox}{\bf Example}
\newtheorem{theorem}{\usebox{\theorembox}}

\def\squarebox#1{\hbox to #1{\hfill\vbox to #1{\vfill}}}

\newcommand{\namedref}[2]{\hyperref[#2]{#1~\ref*{#2}}}

\newcommand{\ignore}[1]{}

\newcommand{\Ceiling}[1]{{{\left \lceil #1 \right \rceil}}}
\newcommand{\Floor}[1]{{{\left \lfloor #1 \right \rfloor}}}

\newcommand{\etal}{{\it et al.}\xspace}

\newcommand{\squishlist}{
   \begin{list}{$\bullet$}
    { \setlength{\itemsep}{0pt}      \setlength{\parsep}{3pt}
      \setlength{\topsep}{3pt}       \setlength{\partopsep}{0pt}
      \setlength{\leftmargin}{1.5em} \setlength{\labelwidth}{1em}
      \setlength{\labelsep}{0.5em} } }

\newcommand{\squishend}{
    \end{list}  }

\newenvironment{mydefinition}[1]{\bigskip \begin{minipage}{0.92\columnwidth}
            \noindent\textsc{Definition} \textsf{({#1})}. \it} {\end{minipage}\bigskip}

\newcommand{\greedy}{{\sc greedy}\xspace}
\newcommand{\lookahead}{{\sc lookahead}\xspace}
\newcommand{\papillon}{Papillon\xspace}

\title{\bf \LARGE Papillon: Greedy Routing in Rings}
\author{
  {\bf Ittai Abraham}\footnote{School of Computer Science and Engineering,
       Hebrew University of Jerusalem, Israel.
       E-Mail: {\tt ittaia@cs.huji.ac.il}}
 \and
  {\bf Dahlia Malkhi}\footnote{Microsoft Research, Silicon Valley and
       School of Computer Science and Engineering,
       Hebrew University of Jerusalem, Israel.
       E-Mail: {\tt dalia@microsoft.com}}
 \and
 {\bf Gurmeet Singh Manku}\footnote{
      Google Inc., USA.
      E-mail: {\tt \small manku@google.com}}
}

\date{}

\begin{document}
\maketitle

 \begin{abstract}
   We study {\sc greedy} routing over $n$ nodes placed in a ring, with
   the  \emph{distance} between two nodes  defined to be the clockwise
   or the absolute distance between them  along the ring.  Such graphs
   arise in the  context of  modeling social  networks and in  routing
   networks for peer-to-peer  systems.  We construct the first network
   over $n$ nodes in which {\sc greedy} routing takes $O(\log n / \log
   d)$ hops in the worst-case, with $d$ out-going  links per node. Our
   result   has  the    first asymptotically   optimal  greedy routing
   complexity.     Previous  constructions  required   $O(\frac{\log^2
     n}{d})$ hops.
 \end{abstract}

% %%%%%%%%%%%%%%%%%%%%%%%%%%%%%%%%%%%%%%%%%%

 \section{Introduction}
 \label{sec:introduction}

 We study  \greedy  routing over uni-dimensional  metrics\footnote{The
 principles  of this work  can  be  extended  to higher  dimensional
 spaces. We focus on one-dimension   for simplicity.}  defined  over
 $n$  nodes lying in   a ring.  \greedy  routing   is the strategy  of
 forwarding a message along  that  out-going edge that minimizes   the
 {\it distance} remaining to the destination:

 \begin{mydefinition}{Greedy Routing}
   In a graph $(V,E)$ with a given distance function $\delta: V \times
   V \rightarrow \mathcal{R}^+$, \greedy routing entails the following
   decision: Given a target node $t$, a node $u$ with neighbors $N(u)$
   forwards  a  message  to its  neighbor   $v  \in  N(u)$ such   that
   $\delta(v,t) = \min_{x \in N(u)} \delta(x,t)$.
 \end{mydefinition}

 \noindent
 Two {\it natural} distance metrics  over $n$ nodes placed in a circle
 are the clockwise-distance and the absolute-distance between pairs of
 nodes:
 \begin{eqnarray*}
   \delta_{clockwise}(u, v) & = &
   \begin{cases}
      v-u   & v\geq u\\
      n+v-u & \text{otherwise}
   \end{cases}\\
   \delta_{absolute}(u, v)  & = &
   \begin{cases}
      \min \{ v-u, n+u-v \} & v\geq u\\
      \min \{ u-v, n+v-u \} & \text{otherwise}
   \end{cases}
 \end{eqnarray*}

 \noindent
 In this paper, we study the following related problems for the
 above distance metrics:

 \begin{center}
 \begin{minipage}{0.95\textwidth} \it
   \squishlist

   \item[I.]  Given integers  $d$ and  $\Delta$, what  is  the largest
      graph that satisfies two constraints: the out-degree of any node
      is at most  $d$, and the length of the  longest \greedy route is
      at most $\Delta$ hops?

   \item[II.] Given  integers $d$ and  $n$, design a network  in which
      each node has out-degree at most $d$ such that the length of the
      longest \greedy route is minimized.

   \squishend
 \end{minipage}
 \end{center}

 \subsection*{Summary of results}

 \begin{enumerate}

 \item  We  construct  a   family  of  network  topologies,  the  {\em
    \papillon}\footnote{Our   constructions   are   variants  of   the
    well-known butterfly family, hence  the name \papillon.}, in which
    {\sc  greedy}   routes  are  asymptotically   optimal.   For  both
    $\delta_{clockwise}$   and   $\delta_{absolute}$,  \papillon   has
    \greedy routes of  length $\Delta = \Theta(\log n  / \log d)$ hops
    in  the  worst-case  when  each  node  has  $d$  out-going  links.
    \papillon is  the first construction  that achieves asymptotically
    optimal worst-case \greedy routes.

 \item Upon further investigation:, two properties of  \papillon emerge:
     (a) \greedy routing does  not send messages along shortest paths,
     and (b)  Edge congestion with \greedy routing  is not  uniform --
     some edges are used more often  than others. We exhibit the first
     property  by identifying routing  strategies that result in paths
     shorter than those  achieved  by {\sc greedy} routing.   In fact,
     one of these strategies guarantees uniform edge-congestion.

 \item Finally, we consider another distance function
   $\delta_{xor}(u,  v)$, defined as   the number of  bit-positions in
   which $u$ and $v$ differ. $\delta_{xor}$ occurs naturally, e.g., in
   hypercubes, and  \greedy routing  with $\delta_{xor}$  routes along
   shortest  paths in them.  We construct  a variant of \papillon that
   supports  asymptotically optimal routes  of length $\Theta(\log n /
   \log  d)$  in  the worst-case,  for  \greedy  routing with distance
   function $\delta_{xor}$.

 \end{enumerate}

% %%%%%%%%%%%%%%%%%%%%%%%%%%%%%%%%%%%%%%%%%%
% %%%%%%%%%%%%%%%%%%%%%%%%%%%%%%%%%%%%%%%%%%

 \section{Related Work}
 \label{sec:related}

 \greedy  routing is  a fundamental  strategy in  network  theory.  It
 enjoys numerous advantages.  It  is completely decentralized, in that
 any node  takes routing decisions  locally and independently.   It is
 oblivious,  thus  message  headers  need  not be  written  along  the
 route. It is inherently fault tolerant, as progress toward the target
 is guaranteed so  long as some links are available.   And it has good
 locality behavior  in that every  step decreases the distance  to the
 target.   Finally,  it  is   simple  to  implement,  yielding  robust
 deployments.   For  these  reasons,  {\sc greedy}  routing  has  long
 attracted  attention in  the research  of network  design.  Recently,
 \greedy routing  has witnessed increased research  interest in the
 context of  decentralized networks.  Such networks  arise in modeling
 social networks  that exhibit the ``small world  phenomenon'', and in
 the design of overlay networks for peer-to-peer (P2P) systems. We now
 summarize known results pertaining to \greedy routing on a circle.

 \subsection*{The Role of the Distance Function}

 Efficient graph constructions are  known that support \greedy routing
 with   distance   function   other  than   $\delta_{clockwise}$,
 $\delta_{absolute}$ and $\delta_{xor}$.
 For de Bruijn networks,  the traditional
 routing algorithm  (which routes almost always  along shortest paths)
 corresponds  to \greedy routing  with $\delta(u,  v)$ defined  as the
 longest suffix of $u$ that is also the prefix of $v$.  For a 2D grid,
 shortest  paths correspond  to  \greedy routing  with $\delta(u,  v)$
 defined  as  the  Manhattan  distance  between  nodes  $u$  and  $v$.

 For \greedy routing on a circle, the best-known constructions have $d
 = \Theta(\log n)$ and $\Delta  =  \Theta(\log n)$. Examples  include:
 Chord~\cite{chord:sigcomm01}           with         distance-function
 $\delta_{clockwise}$,  a   variant   of   Chord  with ``bidirectional
 links''~\cite{ganesan:soda04}           and         distance-function
 $\delta_{absolute}$, and   the   hypercube  with    distance function
 $\delta_{xor}$. In this paper, we improve upon all of these
 constructions by showing how to route in
 $\Theta(\log n / \log d)$ hops in the worst case with
 $d$ links per node.

 \subsection*{\greedy Routing in Deterministic Graphs}

 The  \textsf{Degree-Diameter  Problem},  studied  in  extremal  graph
 theory, seeks  to identify the largest graph  with diameter $\Delta$,
 with each node having  out-degree at most $d$ (see Delorme~\cite{ddp}
 for a survey).  The best  constructions for large $\Delta$ tend to be
 sophisticated~\cite{bermond:92,comellas:92,exoo:01}.    A  well-known
 upper bound  is $N(d, \Delta)  = 1 +  d + d^2  + \cdots +  d^\Delta =
 \frac{d^{\Delta+1}  - 1}{d-1}$,  also known  as the  Moore  bound.  A
 general lower  bound is $d^\Delta + d^{\Delta-1}$,  achieved by Kautz
 digraphs~\cite{kautz:68,kautz:69}, which are  slightly superior to de
 Bruijn  graphs~\cite{debruijn:46} whose  size is  only
 $d^\Delta$. Thus  it is   possible to route
 in $O(\log  n / \log  d)$ hops in  the worst-case with  $d$ out-going
 links  per  node.  Whether  \greedy  routes  with distance  functions
 $\delta_{clockwise}$  or  $\delta_{absolute}$  can achieve  the  same
 bound, is the question we have addressed in this paper.

 \greedy routing  with distance function  $\delta_{absolute}$ has been
 studied for  Chord~\cite{ganesan:soda04}, a popular  topology for P2P
 networks.  Chord  has $2^b$ nodes,  with out-degree $2b-1$  per node.
 The longest \greedy route takes  $\Floor{b/2}$ hops.  In terms of $d$
 and $\Delta$, the  largest-sized Chord network has $n  = 2^{2\Delta +
 1}$ nodes.  Moreover, $d$ and $\Delta$ cannot be chosen independently
 -- they  are  functionally  related.    Both  $d$  and  $\Delta$  are
 $\Theta(\log n)$.   Analysis of \greedy routing of  Chord leaves open
 the  following question:

 \smallskip
 \centerline{\it For  \greedy routing  on a  circle, is
 $\Delta = \Omega(\log n)$ when $d = O(\log n)$?}
 \smallskip

 Xu \etal~\cite{xu:infocom03} provide a partial answer to the above question
 by   studying   \greedy   routing  with   distance   function
 $\delta_{clockwise}$ over  \emph{uniform} graph topologies.   A graph
 over $n$ nodes placed in a circle is said to be uniform if the set of
 clockwise   offsets  of   out-going  links   is  identical   for  all
 nodes. Chord  is an example of  a uniform graph.  Xu  \etal show that
 for  any uniform  graph  with  $O(\log n)$  links  per node,  \greedy
 routing  with  distance  function  $\delta_{clockwise}$  necessitates
 $\Omega(\log n)$ hops in the worst-case.

 Cordasco \etal~\cite{fchord:sirocco04} extend the result of Xu
 \etal~\cite{xu:infocom03}  by  showing   that  \greedy  routing  with
 distance function  $\delta_{clockwise}$ in  a uniform graph  over $n$
 nodes satisfies the inequality $n \leq  F(d + \Delta + 1)$, where $d$
 denotes the  out-degree of each node,  $\Delta$ is the  length of the
 longest  \greedy  path, and  $F(k)$  denotes  the $k^{th}$  Fibonacci
 number.  It  is well-known that  $F(k) = [\phi^k /  \sqrt{5}]$, where
 $\phi  = 1.618\ldots$  is  the  Golden ratio  and  $[x]$ denotes  the
 integer closest to  real number $x$.  It follows  that $1.44 \log_2 n
 \leq d  + \Delta  + 1$.  Cordasco \etal show  that the  inequality is
 strict  if $|d  -  \Delta| >  1$. For  $|d  - \Delta|  \leq 1$,  they
 construct uniform  graphs based upon Fibonacci  numbers which achieve
 an optimal tradeoff between $d$ and $\Delta$.

 \medskip

 The  results  in ~\cite{ganesan:soda04,xu:infocom03,fchord:sirocco04}
 leave open  the question whether there exists  any graph construction
 that permits \greedy routes of  length $\Theta(\log n / \log d)$ with
 distance  function  $\delta_{clockwise}$ and/or  $\delta_{absolute}$.
 \papillon  provides  an  answer  to  the problem  by  constructing  a
 non-uniform graph --- the set of clockwise offsets of out-going links
 is different for different nodes.

 \subsection*{\greedy Routing in Randomized Graphs}

 \greedy  routing  over  nodes  arranged  in a  ring  with  distance
 function $\delta_{clockwise}$  has recently been  studied for certain
 classes of {\it randomized} graph constructions. Such graphs arise in
 modeling social networks that exhibit the ``small world phenomenon'',
 and in the design of overlay networks for P2P systems.

 In   the  seminal  work   of  Kleinberg   \cite{kleinberg:stoc00},  a
 randomized  graph was  constructed in  order to  explain  the ``small
 world         phenomenon'',        first         identified        by
 Milgram~\cite{milgram:pt67}. The phenomenon refers to the observation
 that individuals are able to  route letters to unknown targets on the
 basis  of knowing  only their  immediate social  contacts.  Kleinberg
 considers  a set  of nodes  on  a uniform  two-dimensional grid.   It
 proposes  a  link  model in  which  each  node  is connected  to  its
 immediate grid  neighbors, and in  addition, has a single  long range
 link drawn  from a normalized  harmonic distribution with  power $2$.
 In the resulting graph, \greedy  routes have length at most $O(\log^2
 n)$ hops in expectation; this  complexity was later shown to be tight
 by Barri{\`e}re \etal in \cite{barriere:disc01}.

 Kleinberg's construction has  found  applications  in the design   of
 overlay  routing   networks    for     Distributed Hash    Tables.
 Symphony~\cite{symphony:usits03}  is   an  adaptation of  Kleinberg's
 construction in a single  dimension.  The idea  is to place $n$ nodes
 in a virtual circle and to equip each  node with $d \geq 1$ out-going
 links.  In the resulting network, the  average path length of \greedy
 routes      with    distance     function  $\delta_{clockwise}$    is
 $O(\frac{1}{d}\log^2 n)$ hops.  Note that unlike Kleinberg's network,
 the space here is virtual and  so are the distances  and the sense of
 \greedy routing.  The same  complexity  was achieved with a  slightly
 different        Kleinberg-style     construction     by       Aspnes
 \etal~\cite{aspnes:podc02}.  In the   same paper, it  was  also shown
 that  any     symmetric,   randomized   degree-$d$      network   has
 $\Omega(\frac{\log^2 n}{d\log\log n})$ \greedy routing complexity.

 Papillon outperforms all of the above randomized constructions, using
 degree $d$ and achieving  $\Theta(\log n/\log d)$ routing.  It should
 be  possible to  randomize Papillon  along similar  principles to the
 Viceroy\cite{viceroy:podc02} randomized construction of the butterfly
 network, though we do not pursue this direction here.

 \subsection*{Summary of Known Results}

 With $\Theta(\log n)$ out-going  links per node, several graphs  over
 $n$ nodes  in a circle support  \greedy routes  with $\Theta(\log n)$
 {\sc greedy} hops.  Deterministic graphs with this property include:
 (a) the original  Chord~\cite{chord:sigcomm01} topology with distance
 function $\delta_{clockwise}$,   (b)  Chord with  edges   treated  as
 bidirectional~\cite{ganesan:soda04}    with    distance      function
 $\delta_{absolute}$.   This is also    the known lower  bound  on any
 uniform    graph  with  distance     function    $\delta_{clockwise}$
 \cite{xu:infocom03}.    Randomized   graphs  with  the  same tradeoff
 include  randomized-Chord~\cite{gummadi:sigcomm03,zhang:sigmetrics03}
 and Symphony~\cite{symphony:usits03}  -- both with  distance function
 $\delta_{clockwise}$.   With   degree   $d  \le   \log n$,   Symphony
 \cite{symphony:usits03} has  \greedy routes of length $\Theta((\log^2
 n)/ d)$    on  average.  The   network  of  \cite{aspnes:podc02} also
 supports \greedy routes of length $O((\log^2 n)/d)$ on average , with
 a   gap   to  the   known  lower    bound   on   their   network   of
 $\Omega(\frac{\log^2 n}{d\log\log n})$.

 The above results are  somewhat discouraging, because routing that is
 \textbf{non}-{\sc  greedy}  can  achieve  much  better  results.   In
 particular, networks  of degree $2$  with hop complexity  $O(\log n)$
 are  well known,  e.g.,  the Butterfly  and  the de  Bruijn (see  for
 example \cite{leighton:92} for exposition material).  And networks of
 logarithmic  degree  can  achieve  $O(\log n/  \log\log  n)$  routing
 complexity (e.g., take the  degree-$\log_2 n$ de Bruijn).  Routing in
 these networks is non-{\sc greedy} according to any one of our
 metrics ($\delta_{clockwise}$, $\delta_{absolute}$, and
 $\delta_{xor}$).

 The \papillon\  construction demonstrates  that we can  indeed design
 networks  in which  {\sc greedy}  routing  along  these metrics   has
 asymptotically optimal  routing  complexity.  Our  contribution  is a
 family  of networks that extends the  Butterfly network family, so as
 to  facilitate  efficient {\sc  greedy} routing.  With  $d$ links per
 node, {\sc    greedy}  routes are  $\Theta(\log  n/\log   d)$  in the
 worst-case, which is asymptotically  optimal.  For  $d = o(\log  n)$,
 this  beats the  lower  bound of  \cite{aspnes:podc02}  on symmetric,
 randomized  greedy routing networks  (and  it meets it for  $d=O(\log
 n$).  In the specific case of $d=\log n$, our greedy routing achieves
 $O(\log n/\log \log n)$ average route length.

\subsection*{\greedy with \lookahead}

 Recent  work~\cite{manku:stoc04} explores the surprising advantages of
 \greedy with   \lookahead in  randomized graphs  over  $n$ nodes  in a
 circle.  The  idea behind \lookahead  is  to take neighbor's neighbors
 into  account to make  routing  decisions.  It  shows that greedy with
\lookahead achieves  $O(\log^2 n/ d \log d)$  expected route length in
Symphony~\cite{symphony:usits03}.    For   other networks  which  have
$\Theta(\log     n)$      out-going   links      per    node,    e.g.,
randomized-Chord~\cite{gummadi:sigcomm03,zhang:sigmetrics03},
randomized-hypercubes~\cite{gummadi:sigcomm03},
skip-graphs~\cite{aspnes:soda03}  and  SkipNet~\cite{skipnet:usits03},
average path length  is $\Theta(\log n  / \log  \log  n)$ hops.  Among
these networks, Symphony and randomized-Chord use \greedy routing with
distance function $\delta_{clockwise}$. Other networks use a different
distance  function  (none of them uses  $\delta_{xor}$).   For each of
these networks,  with  $O(\log n)$ out-going   links per node,  it was
established   that   plain  \greedy   (\emph{without} \lookahead)   is
sub-optimal and achieves $\Omega(\log n)$ expected route lengths.  The
results suggest that {\sc lookahead} has significant impact on \greedy
 routing.

 Unfortunately,  realizing    \greedy routing   with   \lookahead on a
 degree-$k$ network implies that $O(k^2)$  nodes need to be considered
 in each hop,  while plain \greedy needs  to consider only $k$ nodes.
 For $k= \log_2 n$, this implies a $O(\log n)$ overhead for \lookahead
 routing in every hop.

 \papillon demonstrates that  it is possible to  construct a graph  in
 which  each node has degree  $d$  and in which \greedy \emph{without}
 1-\lookahead has routes of  length $\Theta(\log n  / \log d)$  in the
 worst case, for the metrics $\delta_{clockwise}$, $\delta_{absolute}$
 and $\delta_{xor}$.  Furthermore, for all $d = o(\log n)$, plain {\sc
   greedy} on  our network design beats  even the  results obtained in
 \cite{manku:stoc04} with $1$-{\sc lookahead}.

 \subsection*{Previous Butterfly-based Constructions}

 \noindent

 Butterfly networks have been used  in the context of routing networks
 for DHTs as follows:
 \begin{enumerate}

 \item Deterministic butterflies have been proposed for DHT routing by
    Xu  \etal~\cite{xu:infocom03},  who  subsequently developed  their
    ideas into  Ulysses~\cite{ulysses:icnp03}.  \papillon for distance
    function  $\delta_{clockwise}$  has  structural similarities  with
    Ulysses -- both are butterfly-based networks.  The key differences
    are as  follows: (a) Ulysses  does not use  $\delta_{absolute}$ as
    its distance  function, (b) Ulysses does not  use \greedy routing,
    and  (c)  Ulysses uses  more  links  than  \papillon for  distance
    function  $\delta_{clockwise}$  --   additional  links  have  been
    introduced  to ameliorate  non-uniform edge  congestion  caused by
    Ulysses' routing algorithm. In contrast, the {\sc congestion-free}
    routing algorithm  developed in \S\ref{sec:improved}  obviates the
    need    for    any   additional    links    in   \papillon    (see
    Theorem~\ref{thm:congestion_free_clockwise}).

 \item Viceroy~\cite{viceroy:podc02}  is a \emph{randomized} butterfly
    network  which routes  in  $O(\log n)$  hops  in expectation  with
    $\Theta(1)$      links      per      node.      Mariposa      (see
    reference~\cite{dipsea:2004} or~\cite{manku:podc03}) improves upon
    Viceroy by providing routes of length  $O(\log n / \log d)$ in the
    worst-case,  with  $d$  out-going  links per  node.   Viceroy  and
    Mariposa are different from  other randomized networks in terms of
    their design philosophy.
 The \papillon\ topology borrows elements of the geometric embedding of the
 butterfly in a circle from Viceroy \cite{viceroy:podc02} and from
 \cite{manku:podc03}, while extending them for {\sc greedy} routing.

 \end{enumerate}

 \section{\papillon}
 \label{sec:papillon}

 We  construct  two variants   of   butterfly networks, one  each  for
 distance-functions $\delta_{clockwise}$ and $\delta_{absolute}$.  The
 network has $n$ nodes arbitrarily positioned  on a ring. We label the
 nodes from $0$ to  $n-1$ according to their  order on the  ring.  For
 convenience,  $x \bmod n$  always represents an  element lying in the
 range $[0, n-1]$ (even when $x$ is negative, or greater than $n-1$).

 \begin{mydefinition}{\papillon for $\delta_{clockwise}$}
   ${\mathcal B}_{clockwise}(\kappa,m)$  is a directed  graph, defined
   for any pair of integers $\kappa,m \geq 1$
   \begin{enumerate}

   \item Let $n =  \kappa^m m$.

   \item  Let $\ell(u) \equiv  (m-1) -  (u \bmod  m)$.  Each  node has
      $\kappa$ links. For node $u$,  these directed links are to nodes
      $(u + x) \bmod n$, where
      $ x \in  \{1 + im\kappa^{\ell(u)}\ |\ i \in [0, \kappa -1 ] \}$.

      We  denote  the  link  with   node  $(u+1)  \bmod  n$  as  $u$'s
      \emph{``short  link''}.  The other  $\kappa-1$ links  are called
      $u$'s \emph{``long links''}.

   \end{enumerate}
 \end{mydefinition}

 \begin{mydefinition}{\papillon for $\delta_{absolute}$}
   ${\mathcal  B}_{absolute}(k,m)$ is a  directed graph,  defined for
   any pair of integers $k,m \geq 1$,
   \begin{enumerate}

   \item Let $n =(2k+1)^m m$.

   \item  Let $\ell(u) \equiv  (m-1) -  (u \bmod  m)$.  Each  node has
      $2k+2$ out-going links.  Node  $u$ makes $2k+1$ links with nodes
      $(u + x) \bmod n$, where
      $ x  \in  \{1  +   im(2k+1)^{\ell(u)}\ |\ i \in [-k, +k] \}$.
      Node $u$ also makes an out-going link with node $(u+x) \bmod n$,
      where $x = -m+1$.

      We denote  the link with node $(u+1) \bmod n$
      as  $u$'s \emph{``short  link''}.   The other  $2k+1$ links  are
      called $u$'s \emph{``long links''}.

    \end{enumerate}
 \end{mydefinition}

 In both ${\mathcal B}_{clockwise}$ and ${\mathcal B}_{absolute}$, all
 out-going links  of node   $u$ are incident    upon nodes with  level
 $(\ell(u) - 1) \bmod  m$.   In ${\mathcal B}_{clockwise}$, the  short
 links   are    such   that each    hop    diminishes  the   remaining
 \emph{clockwise} distance by at least one. Therefore, \greedy routing
 is guaranteed   to take  a finite   number  of  hops.   In ${\mathcal
   B}_{absolute}$, not  every  \greedy  hop  diminishes the  remaining
 \emph{absolute} distance.  However,  \greedy routes are  still finite
 in length, as we show in the proof of Theorem~\ref{thm:absolute}.

 \begin{theorem} \label{thm:clockwise}
   \greedy routing in ${\mathcal  B}_{clockwise}$ with distance
   function $\delta_{clockwise}$ takes $3m-2$ hops in
   the worst-case.  The average is less than $2m-1$ hops.
 \end{theorem}

 \begin{proof}
  For any node $u$, we define
  $
  \text{SPAN}(u)\equiv \{  v \  |\ 0  \leq \delta_{clockwise}(u,  v)  < m
  \kappa^{\ell(u)+1} \}.
  $
  Let  $t$ and $u$ denote the target node and the current node,
  respectively. Routing
  proceeds in (at most) three phases:
  \begin{center}
    \begin{tabular}{lll}
      Phase I: & $t \not\in \text{SPAN}(u)$ & (at most $m-1$ hops)\\
      Phase II: & $t \in \text{SPAN}(u)$ and
            $\delta_{clockwise}(u,t)  \ge  m$ & (at most $m$ hops)\\
      Phase III: & $t \in \text{SPAN}(u)$ and
            $\delta_{clockwise}(u,t) <  m$ & (at most $m-1$ hops)
    \end{tabular}
  \end{center}
  We now prove upper bounds on the number of hops in each phase.
  \begin{enumerate}

  \item[I.]
     The out-going  links of $u$ are  incident upon nodes at
     level $(\ell(u) - 1) \bmod  m$. So eventually, the level of the
     current node
     $u$ will be $m-1$. At  this point,
     $t \in \text{SPAN}(u)$ because $\text{SPAN}(u)$ includes
     \emph{all} the
     nodes.  Thus Phase 1 lasts for at most $m-1$ hops
     ($\frac{m-1}{2}$ hops on
     average).

  \item[II.]
     \greedy will  forward  the
     message  to some node  $v$  such that  $t  \in \text{SPAN}(v)$  and
     $\ell(v)=\ell(u)-1$.  Eventually, the current node $u$ will
     satisfy the property   $\ell(u) = 0$.  This node will  forward
     the message  to some node
     $v$ with $\ell(v) = m-1$ such that $\delta_{clockwise}(v,t) < m$,
     thereby terminating this phase of  routing. There are at most $m$
     hops in this phase (at most $m$ on average as well).

  \item[III.]
     In this phase, \greedy  will  decrease  the clockwise
     distance by  exactly one in each hop by following the
     short-links. Eventually, target $t$ will be reached. This phase
     takes  at  most  $m-1$  hops  ($\frac{m-1}{2}$  hops  on
     average).
  \end{enumerate}

  The worst-case  route length is  $3m-2$.
  On average, routes are at most $2m-1$ hops long.
 \end{proof}

 \begin{theorem} \label{thm:absolute}
   \greedy routing  in ${\mathcal B}_{absolute}$ with distance
   function $\delta_{absolute}$ takes  $3m-2$ hops in
   the worst-case.  The average is less than $2m-1$ hops.
 \end{theorem}
 \begin{proof}
  For any node $u$, we define
  \[
  \text{SPAN}(u)\equiv  \{ v \ |\  \delta_{absolute}(u, v) =  | c +
  m\sum_{i=0}^{\ell(u)} (2k+1)^i d_i |,\ c \in [0, m-1],\ d_i \in [-k,
  +k] \,\}.
  \]

  Let  $t$ and $u$ denote the target node and the current node,
  respectively.

  Routing  proceeds in (at most) three phases:
  \begin{center}
    \begin{tabular}{lll}
      Phase I: & $t \not\in \text{SPAN}(u)$ & (at most $m-1$ hops)\\
      Phase II: & $t \in \text{SPAN}(u)$ and
            $\delta_{absolute}(u,t)  \ge  m$ & (at most $m$ hops)\\
      Phase III: & $t \in \text{SPAN}(u)$ and
            $\delta_{absolute}(u,t) <  m$ & (at most $m-1$ hops)
    \end{tabular}
  \end{center}
  We now prove upper bounds on the number of hops in each phase.

  \begin{enumerate}

  \item[I.] All out-going  links of node $u$ are  incident upon nodes at
     level $(\ell(u) - 1) \bmod  m$. So eventually, the current node
     $u$ will satisfy the property $\ell(u) = m - 1$. At this point,
     $t \in \text{SPAN}(u)$ because $\text{SPAN}(u)$ includes
     \emph{all} nodes. Thus Phase I lasts at most  $m-1$ hops
     (at most $\frac{m-1}{2}$ hops on average).

  \item[II.]
     Phase  2 terminates if target node $t$  is reached, or if
     $\delta_{absolute}(u, t)  < m$.
      Node
     $u$ always forwards the  message  to some  node  $v$ such that
     $t \in \text{SPAN}(v)$  and  $\ell(v)  = \ell(u)  -  1$. So
     eventually, either target $t$ is reached, or the current node
     $u$ satisfies the property
     $\ell(u) =  0$.  At this point, if node $u$  forwards the  message to
      node $v$, then it is guaranteed that $\ell(v) =
     m-1$ and  $\delta_{absolute}(v,t) <  m$,
     thereby terminating Phase II. There are at most $m$
     hops in this phase (at most $m$ on average as well).

  \item[III.]     The target node $t$ is reached in at most $m-1$
    hops (the existence of  the ``back
     edge''  that connects node  $u$ to  node $(u  + 1  - m)  \bmod n$
     guarantees this).  This  phase takes at most $m-1$  hops (at most
     $\frac{m-1}{2}$ hops on average).
  \end{enumerate}

  The worst-case  route length is  $3m-2$.
  On average, routes are at most $2m-1$ hops long.
 \end{proof}

 \bigskip

 Routes  in both ${\mathcal  B}_{clockwise}$ and  ${\mathcal B}_{absolute}$
 are  at  most $3m-2$  hops,  which is  $O(\log  (\kappa^m  m) /  \log
 \kappa)$  and $O(\log  ((2k+1)^m  m) /  \log (2k+2))$,  respectively.
 Given degree $d$ and diameter $\Delta$, the size of \papillon is $n =
 2^{O(\Delta)}\Delta$ nodes.   Given degree $d$ and  network size $n$,
 the longest route has length $\Delta = O(\log n / \log d)$.

% %%%%%%%%%%%%%%%%%%%%%%%%%%%%%%%%%%%%%%%%%%
% %%%%%%%%%%%%%%%%%%%%%%%%%%%%%%%%%%%%%%%%%%

 \section{Improved Routing Algorithms for \papillon}
 \label{sec:improved}

 \greedy  routing does  not route  along shortest-paths  in ${\mathcal
 B}_{clockwise}$ and  ${\mathcal B}_{absolute}$.  We  demonstrate this
 constructively below,  where we study a routing  strategy called {\sc
 hypercubic-routing} which achieves shorter path lengths than \greedy.

 \subsection*{Hypercubic Routing}

 \begin{theorem} \label{thm:fast_clockwise}
   There exists  a routing strategy for  ${\mathcal B}_{clockwise}$ in
   which routes take $2m-1$ hops  in the worst-case. The average is at
   most $1.5m$ hops.
 \end{theorem}

 \begin{proof}
   Consider  the  following   {\sc  hypercubic-routing}  algorithm  on
   ${\mathcal B}_{clockwise}$.   Let $s$ be  the source node,  $t$ the
   target,  and  let  $dist =  \delta_{clockwise}(s,  t)  =  c +  m  +
   m\sum_{i=0}^{i=m-1} \kappa^i  d_i$ with $0  \leq c <m$ and  $0 \leq
   d_i <  \kappa$ ($dist$ has exactly one  such representation, unless
   $dist \leq m$ in which case routing takes $< m$ hops).

   Phase I: Follow  the short-links to ``fix'' the  $c$-value to zero.
   This takes at most $m-1$ hops (at most $0.5m$ hops on average).

   Phase II: In exactly $m$ hops, ``fix'' the $d_i$'s in succession to
   make  them  all  zeros:  When  the  current node  is  $u$,  we  fix
   $d_{\ell(u)}$ to zero by following the appropriate long-link, i.e.,
   by    shrinking   the    clockwise    distance   by    $d_{\ell(u)}
   \kappa^{\ell(u)} m  + 1$.   The new node  $v$ satisfies  $\ell(v) =
   (\ell(u)+m-1) (\bmod~m)$.  When each $d_i$ is zero, we have reached
   the target.

   Overall,  the worst-case  route  length is  $2m-1$.  Average  route
   length is at most $1.5m$.
 \end{proof}

 \begin{theorem} \label{thm:fast_absolute}
   There exists  a routing  strategy for ${\mathcal  B}_{absolute}$ in
   which routes take $2m-1$ hops  in the worst-case. The average is at
   most $1.5m$ hops.
 \end{theorem}
 \begin{proof}
   Let $s$  be the  source node,  $t$ the   target.

   Phase  I: Follow  the short-links  in the  clockwise  direction, to
   reach a  node $s'$ such that  $\ell(s') = \ell(t)$.   This takes at
   most $m-1$  hops (at most  $0.5m$ hops on average).   The remaining
   distance can be expressed as $m + m\sum_{i=0}^{i=m-1} (2k+1)^i d_i$
   where $-k \leq d_i \leq k$. There is a unique such representation.

   Phase II: In exactly $m$ hops, ``fix'' the $d_i$'s in succession to
   make  them  all  zeros:  When  the  current node  is  $u$,  we  fix
   $d_{\ell(u)}$ by following the appropriate long-link, i.e.,
   by traveling  distance $1  + d_{\ell(u)} (2k+1)^{\ell(u)}  m$ along
   the circle  (this distance is positive or  negative, depending upon
   the sign of $d_{\ell(u)}$).  The  new node $v$ satisfies $\ell(v) =
   (\ell(u)-1) (\bmod~m)$.   When each $d_i$ is zero,  we have reached
   the target.

   Overall,  the worst-case  route  length is  $2m-1$.  Average  route
   length is at most $1.5m$.
 \end{proof}

 \bigskip

 Note that the  edges that connect node $u$ to  node $(u+1-m) \bmod n$
 are redundant for {\sc hypercubic-routing} since they are never used.
 However,  these edges  play  a  crucial role  in  \greedy routing  in
 ${\mathcal  B}_{absolute}$ (to  guide the  message to  the  target in
 Phase 3).

 \subsection*{Congestion-Free Routing}

 Theorems~\ref{thm:fast_clockwise}  and ~\ref{thm:fast_absolute} prove
 that \greedy  routing is sub-optimal  in the constants.  {\sc
 hypercubic-routing}, as described above, is faster than
 \greedy. However, it causes {\it edge-congestion}
 because short-links  are used more often than  long-links.  Let $\pi$
 denote the ratio of maximum and  minimum loads on edges caused by all
 $n  \choose   2$  pairwise  routes.    {\sc  hypercubic-routing}  for
 ${\mathcal  B}_{clockwise}$  consists of  two  phases  (see Proof  of
 Theorem~\ref{thm:fast_clockwise}).   The  load  due  to Phase  II  is
 uniform  -- all  edges  (both short-links  and  long-links) are  used
 equally.  However, Phase  I uses only short-links, due  to which $\pi
 \not= 1$.  We now modify the routing scheme slightly to obtain $\pi =
 1$ for both ${\mathcal B}_{clockwise}$ and ${\mathcal B}_{absolute}$.

 \begin{theorem} \label{thm:congestion_free_clockwise}
   There  exists  a  congestion-free  routing strategy  in  ${\mathcal
   B}_{clockwise}$ that takes  $2m - 1$ hops in  the worst-case and at
   most $1.5m$ hops on average, in which $\pi = 1$.
 \end{theorem}
 \begin{proof}
   The  theorem is proved  constructively, by  building a  new routing
   strategy  called {\sc  congestion-free}. This  routing  strategy is
   exactly the same as {\sc hypercubic-routing}, with a small change.

   Let $s$ be the source node,  $t$ the target. Let $c = (t+m-s) \bmod
   m$, the difference in levels between $\ell(s)$ and $\ell(t)$.

   Phase I: For $c$ steps, follow any out-going link, chosen uniformly
   at  random.   We thus  reach  a node  $s'$  such  that $\ell(s')  =
   \ell(t)$.

   Phase II: The remaining  distance is $dist = \delta_{clockwise}(s',
   t)  = m+  m\sum_{i=0}^{i=m-1}  \kappa^i  d_i$ with  $0  \leq d_i  <
   \kappa$.  Continue  with Phase  II of the  {\sc hypercubic-routing}
   algorithm       for      ${\mathcal       B}_{clockwise}$      (see
   Theorem~\ref{thm:fast_clockwise}).

   It is easy to see that in this case, all outgoing links (short- and
   long-)  are used with  equal probability  along the  route.  Hence,
   $\pi = 1$.
 \end{proof}

 \begin{theorem} \label{thm:congestion_free_absolute}
   There  exists  a  congestion-free  routing strategy  in  ${\mathcal
   B}_{absolute}$ that  takes $2m - 1$  hops in the  worst-case and at
   most $1.5m$ hops on average, in which $\pi = 1$.
 \end{theorem}
 \begin{proof}
   We will  ignore the  edges that connect  node $u$ to  node $(u+1-m)
   \bmod  n$   (recall  that  these   edges  are  not  used   in  {\sc
   hypercubic-routing}                   described                  in
   Theorem~\ref{thm:fast_absolute}). We will ensure  $\pi = 1$ for the
   remainder of the edges.

   {\sc  congestion-free} routing follows  the same  idea as  that for
   ${\mathcal                                           B}_{clockwise}$
   (Theorem~\ref{thm:congestion_free_clockwise}):   Let  $s$   be  the
   source  node, $t$  the  target.  Let  $c  = (t+m-s)  \bmod m$,  the
   difference in levels between  $\ell(s)$ and $\ell(t)$.  In Phase I,
   for $c$  steps, we follow  any out-going link, chosen  uniformly at
   random.  We thus reach a  node $s'$ such that $\ell(s') = \ell(t)$.
   In  Phase   II,  we   continue  as  per   Phase  II  of   the  {\sc
   hypercubic-routing}   algorithm   for   ${\mathcal   B}_{absolute}$
   (Theorem~\ref{thm:fast_absolute}).

   \medskip

   An alternate {\sc congestion-free} routing algorithm for ${\mathcal
   B}_{absolute}$  that  routes deterministically  is  based upon  the
   following idea: We express any integer  $a \in [-k, +k]$ as the sum
   of   two   integers:   $a'    =   \Floor{(k+a)/2}$   and   $a''   =
   -\Floor{(k-a)/2}$. It is  easy to verify that $a =  a' + a''$.  Now
   if we list all pairs $\langle a', a''\rangle$ for $a \in [-k, +k]$,
   then each integer in the  range $[-k, +k]$ appears exactly twice as
   a member of some pair.

   Let $s$ be the source node, $t$ the target.  Let $c = (t+m-s) \bmod
   m$, the difference in  levels between $\ell(s)$ and $\ell(t)$.  The
   remaining distance  is $dist = c +  m+ m\sum_{i=0}^{i=m-1} (2k+1)^i
   d_i$ with $-k \leq d_i \leq  k$ (there is a unique way to represent
   $dist$ in this fashion).

   Phase I: For $c$ steps, if  the current node is $u$, then we follow
   the  edge  corresponding to  $d_{\ell(u)}'$,  i.e.,  the edge  that
   covers  distance   $1  +  md_{\ell(u)}'(2k+1)^{\ell(u)}$   (in  the
   clockwise or the anti-clockwise  direction, depending upon the sign
   of $d_{\ell(u)}'$). At the end of  this phase, we reach a node $s'$
   such that $\ell(s') = \ell(t)$.

   Phase II:  Continue with Phase  II of the  {\sc hypercubic-routing}
   algorithm           for          ${\mathcal          B}_{absolute}$
   (Theorem~\ref{thm:fast_absolute}), for exactly $m$ steps.

   Due to the  decomposition of integers in $[-k,  +k]$ into pairs, as
   defined  above, all  outgoing  links (short-  and  long-) are  used
   equally.  Hence, $\pi = 1$.
 \end{proof}

 \bigskip

 {\bf Notes}: In the context  of the current Internet, out-going links
 correspond to full-duplex TCP connections.  Therefore, the undirected
 graph corresponding to ${\mathcal  B}_{absolute}$ is of interest.  In
 this  undirected  graph, it  is  possible  to devise  congestion-free
 routing with  $\pi =  1$, maximum path  length $m +  \Floor{m/2}$ and
 average route-length at most $1.25m$.   This is achieved by making at
 most $\Floor{m/2}$ initial random steps  either in the down or the up
 direction, whichever gets to a node with level $\ell(t)$ faster.

% %%%%%%%%%%%%%%%%%%%%%%%%%%%%%%%%%%%%%%%%%%
% %%%%%%%%%%%%%%%%%%%%%%%%%%%%%%%%%%%%%%%%%%

 \section{\papillon with Distance Function $\delta_{xor}$}
 \label{sec:xor}

 In this Section, we define a variant of \papillon in which \greedy
 routing with distance function $\delta_{xor}$ results in worst-case
 route length $\Theta(\log n / \log d)$, with $n$ nodes, each having
 $d$ out-going links. For integers $s$ and $t$, $\delta_{xor}(s, t)$
 is defined as the number of bit-positions in which the binary
 representations of $s$ and $t$ differ.

 \begin{mydefinition}{\papillon for $\delta_{xor}$}
   ${\mathcal B}_{xor}(\lambda, m)$  is a directed  graph, defined
   for any pair of integers $\lambda, m \geq 1$ where $\lambda$ is a
   power of two.
   \begin{enumerate}

   \item The  network has $n = m\lambda^m$ nodes labeled  from $0$ to
      $n-1$.

   \item Let $u$ denote a node. Let $\ell(u)$ denote the unique
     integer $x \in [0, m-1]$ that satisfies $x\lambda^m \leq u <
     (x+1)\lambda^m$. The node $u$ makes links with nodes with labels
     \[
      ((\ell(u) + 1) \bmod m)\lambda^m + i\lambda^{\ell(u)},
      \quad \mathrm{where}\ \
      0 \leq i < \lambda.
     \]
   Thus, if $(u, v)$ is an edge, then $\ell(v) =
   (\ell(u) + 1) \bmod m$.
   \end{enumerate}
 \end{mydefinition}

 \begin{theorem} \label{thm:xor}
   \greedy routing  in ${\mathcal B}_{xor}$ with distance function
   $\delta_{xor}$ takes  $2m-1$ hops in
   the worst-case.  The average is at most $1.5m$ hops.
 \end{theorem}
 \begin{proof}
   Let the current node be $s$. Let $t$ denote the target node.
   Then $s \oplus t$, the bit-wise exclusive-OR of $s$ and $t$, can
   uniquely be expressed as $c + \sum_{i=0}^{i=m-1}
   \lambda^i d_i$, where $c \geq 0$ and $0 \leq d_i < \lambda$.
   Routing proceeds in two phases. In Phase I, each of the $d_i$ is
   set to zero. This takes at most $m$ steps (at most $m$ on
   average). In Phase II, the most significant
   $\Ceiling{\log_2 m}$ bits of $s \oplus t$ are set to zero,
   thereby
   reaching the target.   This phase takes at most $m-1$ hops (at most
   $\frac{m-1}{2}$ on average).
 \end{proof}

% %%%%%%%%%%%%%%%%%%%%%%%%%%%%%%%%%%%%%%%%%%
% %%%%%%%%%%%%%%%%%%%%%%%%%%%%%%%%%%%%%%%%%%

 \section{Summary}
 \label{summary}

 We presented \papillon, a  variant of  multi-butterfly networks  which supports
 asymptotically optimal \greedy routes of  length $O(\log n / \log d)$
 with distance functions $\delta_{clockwise}$, $\delta_{absolute}$ and
 $\delta_{xor}$,
 when each  node makes  $d$ out-going links,  in an  $n$-node network.
 \papillon  is the  first construction  with this  property.

 \medskip

 Some questions that remain unanswered:
 \begin{enumerate}

 \item {\it  Is it possible to  devise graphs in  which \greedy routes
    with      distance      function     $\delta_{clockwise}$      and
    $\delta_{absolute}$  are  along  shortest-paths?  }   As  Theorems
    ~\ref{thm:fast_clockwise} and ~\ref{thm:fast_absolute} illustrate,
    \greedy routing  on \papillon  do not route  along shortest-paths.
    Is this property inherent in \greedy routes?

 \item {\it What is the  upper-bound for the Problem of Greedy Routing
    on  the Circle?  }   \papillon furnishes  a lower-bound,  which is
    asymptotically     optimal.      However,     constructing     the
    largest-possible graph  with degree $d$ and  diameter $\Delta$, is
    still an interesting combinatorial problem.

 \end{enumerate}

\newcommand{\etalchar}[1]{$^{#1}$}

 \end{document}